\documentstyle[12pt]{article}
\def\hbeta{{{\beta} \over 2}}

\begin{document}

\rightline{CERN-TH.7497/94}
\rightline{arch-ive/9411332}

\begin{center}

{\large\bf Semi-analytical and Monte Carlo results for the
production of four fermions in {\boldmath $e^+ e^-$} collisions}
\vskip 4pt
{\large }

\end{center}

\noindent
Guido MONTAGNA$^a$, Oreste NICROSINI$^{b,}\footnote{\footnotesize
On leave from INFN, Sezione di Pavia, Italy.}$, Giampiero PASSARINO$^c$
and Fulvio PICCININI$^d$ \\

\noindent
$^a$ INFN, Sezione di Pavia,  Italy \\
$^b$ CERN, TH Division, Geneva, Switzerland  \\
$^c$ Dipartimento di Fisica Teorica, Universit\`a di Torino and INFN,
Sezione di Torino, Turin, Italy \\
$^d$ Dipartimento di Fisica Nucleare e Teorica,
Universit\`a di Pavia, and INFN, Sezione di Pavia, Italy \\

\begin{abstract}

{\small In view of the forthcoming experiments at high energy $e^+ e^-$
colliders, LEP2 and beyond, the process $e^+ e^- \to W^+ W^- \to 4f$ has
been investigated for {\it off-shell} $W$'s. After deriving the matrix element
for the process within the helicity amplitude formalism, QED radiative
corrections to the initial state leptons have been included at the leading
logarithmic level in the framework of the electron structure functions. Cross
sections and distributions are computed by means of both
a semi-analytical and a Monte Carlo code, allowing
for several cuts of experimental interest. A Monte Carlo
event generator has also been built, for simulation purposes. Some
illustrative numerical results are included and shortly described. }

\end{abstract}

\begin{center}
Submitted to Physics Letters B
\end{center}

\vfil
\leftline{CERN-TH.7497/94}
\leftline{November 1994}

\eject
In the context of the experiments planned at $e^+ e^-$ colliders such as
LEP2 and
beyond, the process $e^+ e^- \, \to \, W^+ W^- \, \to \, 4\, f$ plays
a special role since it will be used for an accurate measurement of the
$W$-mass and hopefully to test the non-Abelian trilinear couplings of the
standard model as well as the effects of non-standard ``anomalous'' couplings
between the gauge bosons. In order to achieve these experimental goals,
accurate predictions for the observables of the reaction
$e^+ e^- \, \to \, W^+ W^- \, \to \, 4\, f$ are mandatory and in particular
the contribution of all radiative corrections should be kept well under
control.

For the above reasons a number of studies devoted to four fermions production
in electron-positron collisions, including the main effects of radiative
corrections, have recently appeared in the literature and they can be roughly
grouped in two main lines,  which are in some sense complementary:

\begin{itemize}

\item semi-analytical approaches~\cite{gentle,bd},

\item Monte Carlo approaches~\cite{exc,wop,wwf}.

\end{itemize}

\noindent
Previous calculations of the higher order QED corrections to $e^+ e^- \to
W^+ W^- $, including the influence of the finite $W$ width, were
performed in~\cite{nz,fkm}.
The complete study of the reaction $e^+ e^- \to 4\,$fermions
clearly represents a multi-step program and the aim of this letter is to
summarize our novel theoretical approach to it, which includes the
description of $e^+ e^- \to W^+W^- \to 4\,$fermions with initial state
electromagnetic radiation in the leading log approximation.
A complete analysis of the {\it background}
processes and of the {\it exact} ${\cal O}(\alpha)$ radiative corrections
are currently under investigation.
Our main goal in this letter will be to present an illustrative sample of the
numerical results obtained so far by means of a semi-analytical and a Monte
Carlo program.
We start with a brief description of the basic ingredients.
\vskip 10pt


We have undertaken a general study of the process $e^+e^- \to 4\,$ fermions
with the aim of producing and discussing a wide number of realistic
distributions. As a first step we briefly introduce some basic concepts
and apply them to the {\it double-resonating} diagrams $e^+e^- \to
W^+W^- \to 4\,$fermions. The main emphasis in our approach is on a
covariant description of the kinematics, which allows us to translate
any sort of kinematical cut directly into an analytical constraint on the
integration variables. As for any $2 \to 4$ process we can introduce $15$
invariants $s_{ij} = -(\epsilon_i\,p_i + \epsilon_j\,p_j)^2$ where
$p_i, i=1,\dots,6$ are all the momenta assumed to flow inwards and
$\epsilon_i = \pm 1$ for incoming (outgoing) particles.
There are $6$ linear and $1$ non-linear
constraints among the $s_{ij}$ with a total of $8$ independent invariants,
one of which is of course the invariant mass of the event, $s$. Using
standard techniques all the components of the momenta (up to a sign ambiguity)
can be expressed in terms of the $s_{ij}$ and thus in terms of the
independent invariants. This is why we can express any kinematical cut as
an analytical constraint on the integration variables and why our
multi-dimensional phase space can be treated exactly.

For $e^+(p_1)e^-(p_2) \to f_1(p_3) + \dots f_4(p_6)$ the independent variables
used to describe the process are, besides $s = -(p_1 + p_2)^2$, the following
ones: $z_1 = x_{13}+x_{23}$, $z_3 = x_{15}+x_{25}$, $\mu_1^2 = x_{34}$,
$\mu_2^2 = x_{56}$, $\mu_3^2 = x_{35}$, $\tau_1 = x_{13}$
and $\tau_2 = x_{14}$ where we have defined $x_{ij}s = s_{ij}$.
In the c.m. system they have a rather simple interpretation, namely

\begin{equation}
z_i = {{2 E_i} \over {\sqrt{s}}}, \qquad i = 1,3 ,
\end{equation}

\noindent
where $E_i$ are the energies of particles 1 and 3, which
correspond to the {\it down} fermion coming from the decay of
$W^-$ and to the {\it up} fermion coming from the decay of $W^+$, respectively;

\begin{equation}
\mu_i^2 = {{s_i} \over s}, \qquad i = 1,2 ,
\end{equation}

\noindent
where $s_1$ and $s_2$ are the invariant masses of $W^+$ and $W^-$
respectively;

\begin{equation}
\tau_i = {{E_i} \over \sqrt{s}} (1 - \cos \vartheta_i^+ ), \qquad i = 1,2 ,
\end{equation}

\noindent
where $\vartheta_i^+$ is the scattering angle of the $i$-th particle with
respect to the incoming positron beam;

\begin{equation}
\mu_3^2 = {{s_{13}} \over s},
\end{equation}

\noindent
where $s_{13}$ is the invariant mass between the particles 1 and 3.
Given the independent variables, the further choice of the solution of
the non-linear constraint allows the reconstruction of the kinematics in the
laboratory frame.
The explicit solution of the kinematics allows us to translate several cuts of
experimental interest in terms of boundaries of the independent variables
phase space; such a translation is of great importance from the point of view
of numerical integration when adopting a semi-analytical approach.

When including initial state radiation, two more independent variables
$x_{1,2}$, which represent the energy fraction carried by the
incoming electron and positron after the radiation,
have to be added to the previous set. In this case the above independent
variables are referred to the c.m. system of the hard scattering
process and in this frame the dependent variables are derived by explicit
solution of the kinematics. By means of a Lorentz boost the whole kinematics
in the laboratory system can finally be reconstructed thus allowing
 the boundaries of
the phase space in the presence of realistic cuts  to be
defined also for the radiative process.
\vskip 10pt


Our matrix element contains for the moment
only the three {\it double-resonant}
diagrams in Born approximation, but there is no limitation of the method
that will prevent us from a straightforward generalization to all the
{\it background} processes. The strategy adopted is to use one of the
many formulations of the so-called helicity method~\cite{hel}, which has
the advantage of working directly with invariants. Roughly speaking, each
diagram can be written as

\begin{equation}
d_i(\lambda,\rho_f,\rho_f') = (2\pi)^4\,i g^4 \eta_i \kappa_i \Delta_i
{\bar v}_{-\lambda}\Gamma_{in} u_{\lambda}
{\bar u}_{\rho_f}\Gamma_f v_{-\rho_f} {\bar u}_{\rho_{f'}}\Gamma_{f'}
v_{-\rho_{f'}}
\end{equation}

\noindent
where $\lambda, \rho_f$ and $\rho_{f'}$ are helicity labels,
$\eta_i$ is the sign of the diagram, $\kappa_i$ stands for a collection
of constants and $\Delta_i$ for the collection of propagators. Finally
$\Gamma_{in}, \Gamma_f$ and $\Gamma_{f'}$ are strings of $\gamma$ matrices.
Once each diagram is rewritten as a trace, we explicitly insert for the
bilinears the corresponding expression (valid up to an overall phase),
as for instance

\begin{equation}
v_{\lambda}(p){\bar u}_{\lambda}(q) = - \frac{1}{2}{1\over {(2p\,q)^{1/2}}}\,
\left(1 - \lambda\,\gamma^5\right) p\llap{$/$} q\llap{$/$}
\end{equation}

\noindent
and the trace operation will convert everything in terms of invariants,
which in turn are related to the integration variables.
According to the above procedure
the three diagrams of the process $e^+ e^- \to W^+ W^- \to 4 \, f$
have been computed by means of
{\tt SCHOONSCHIP}~\cite{schoon} and FORTRAN-coded to
get the squared matrix element numerically . The effect of the
electromagnetic initial state radiation is treated in the language of the
QED structure functions~\cite{sf,sf1},
where the corrected cross section can be written in the following form:

\begin{eqnarray}
\sigma (s) = \int d x_1 \, d x_2 \, D(x_1,s) D(x_2,s)
d [PS] {{d \sigma} \over {d [PS]}} .
\end{eqnarray}

\noindent
In the latter equation $d [PS]$ denotes the volume element in the
7-dimensional phase space connected to the independent variables of the
hard scattering reaction in the c.m. frame; $D(x,Q^2)$
provides the probability
of finding, inside a parent electron, an electron of energy fraction $x$ with
virtualness $Q^2$. The explicit expression which we adopt to treat initial
state radiation is given by

\begin{eqnarray}
D(x,s)&=&\, { {\exp \left\{ \frac{1}{2}\beta \,\left( \frac{3}{4} -\gamma_E
\right) \right\} }
\over {\Gamma \left( 1 + \frac{1}{2} \beta \right) } } \,
\hbeta (1 - x)^{\hbeta - 1} - {{\eta} \over 4} (1+x) \\ \nonumber
&+& {1 \over {32}} \eta^2 \left[ -4 (1+x) \ln(1-x) + 3 (1+x) \ln x -
4 {{\ln x} \over {1-x}} - 5 - x \right] ,
\end{eqnarray}

\noindent
with

\begin{eqnarray}
\beta = 2\,{\alpha \over \pi}\,(L-1) , \quad \quad \quad
\eta =  2\,{\alpha \over \pi}\,L  ,
\end{eqnarray}

\noindent
where $L= \ln \left( {s / {m^2}} \right)$ is the collinear logarithm,
$\gamma_E$ is the Euler constant and $\Gamma(z)$ is the gamma function.
The first exponentiated term of Gribov--Lipatov form
describes multiphoton soft radiation, the second and third ones hard
collinear bremsstrahlung. In the structure function the
arbitrary scale $Q^2$ is set equal to $s$ and the hard photon terms are taken
with the coefficient $\eta$ instead of $\beta$
in order to treat the hard radiation in leading log approximation (in this
approximation,  problems connected with gauge invariance are avoided).
\vskip 10pt


As a consequence of the high dimensionality of the integration,
a Monte Carlo strategy could be in principle
competitive with a deterministic numerical
procedure, as far as the balance between number of calls and accuracy are
concerned. Both of the approaches have been adopted to cross-check and put
under control the numerical results. The Monte Carlo integration
(based on the random numbers generator {\tt RANLUX}~\cite{rlux}) has been
performed according to the importance sampling technique in order to cure
the peaking behaviour of the integrand.
This strategy has been helpful also for
the semi-analytical approach, which has been carried out by means of the
NAG~\cite{nag} routine {\tt D01GDF}.
As a result, the deterministic procedure turns out to be more efficient with
regard to a totally inclusive set-up and in the presence of cuts on the
final state invariant masses and/or energies and/or scattering angles of the
outgoing fermions.
On the other hand, the Monte Carlo allows for a  complete and more
flexible control of any kind of experimental cut. Furthermore, the Monte
Carlo routine can also be used as an event generator for simulation purposes.
In this case the events, defined as the components of the four final state
particles momenta, plus the radiative variables
$x_1$ and $x_2$, plus $\sqrt s$, are stored into
proper $n$-tuples. A detailed description of the formulation and
of the technical solutions adopted
will be presented elsewhere.

In the following we show some illustrative  results obtained
by means of the two numerical procedures described above. Our set of input
parameters is $M_{Z} = 91.1887\,$GeV, $M_{W} = 80.22\,$GeV, $\Gamma_{Z}
= 2.4974\,$GeV, $\Gamma_{W} = 2.08\,$GeV~\cite{pdg}.

In Fig.~1 the total cross section, for $W$ decaying in all possible
final states (excluding top production), in Born approximation (dashed line)
is compared with the initial state QED-corrected one (solid line)
as a function of the c.m. energy for a fully
extrapolated set-up. Concerning the tree level
prediction, the dashed line result has been obtained by a two-fold integration
of the formula given in ref.~\cite{jap} while the solid line derives from
a nine-fold integration of the $e^+ e^- \to
W^+ W^- \to 4f$ matrix element performed by our semi-analytical code.
The Monte Carlo results are represented by the open circles,
showing a fully satisfactory agreement in both
the Born and QED-corrected case. For such a configuration,
the effect of initial state radiation is
to reduce the peak cross section by about 10\%, to slightly
shift the peak position towards
 higher energies and to determine the appearance of a radiative tail
as a typical convolution effect. For the QED-corrected total cross section
we compared our results, obtained by properly adapted structure functions,
with the values quoted in~\cite{gentle} and found very good agreement.

Figure~2 shows the $W^-$ invariant mass distribution $d \sigma / d M_-$
($M_-^2 = - (q_1 + q_2)^2$) at the Born level for
 two c.m. energies in the LEP2 regime, namely $\sqrt{s} = 175$~GeV
(dotted line) and $\sqrt{s} = 190$~GeV (dashed line).
In order to single out the contribution of the initial state radiation,
the corrected invariant mass distribution at
 $\sqrt{s} = 175$~GeV (solid line) is also plotted for comparison.
As in Fig.~1, the Monte Carlo
predictions given by the markers fully agree with the semi-analytical results
 used to produce the continuous lines.

The effect of some realistic experimental cuts is shown for
the total cross section (Fig.~3) and the invariant mass distribution
at $\sqrt{s} = 175$~GeV (Fig.~4)
of $e^+ e^- \to W^+ W^- \to 4f$ where both $W$'s are
required to decay hadronically. All the curves include the effect of
the initial state leading log QED corrections. \
The experimental conditions considered are the
following: no cut (solid line), cuts on the energies of all the four final
state fermions $E_{min}^i = 25$~GeV (dashed line), constraint on the invariant
masses of both $W$'s
$77 \leq M_{\pm} \leq 83$~GeV for the total cross section and on the
$W^+$ invariant mass $77 \leq M_{+} \leq 83$~GeV for the
$W^-$ invariant mass distribution (close-dotted line),
acceptance cuts on the scattering angles of the produced fermions
$40^\circ \leq \vartheta_i \leq 140^\circ$ (wide-dotted line). Also for such
realistic configurations the agreement between the
semi-analytical (lines) and
 Monte Carlo (markers)
integrations is fully satisfactory pointing out the flexibility
 of our semi-analytical code.

In Figs.~5--9 we show some distributions of experimental interest
for LEP2 physics, which have been obtained processing the $n$-tuple
created by the event generator at $\sqrt{s} = 175$~GeV taking into
account the emission of initial state radiation. They refer to a sample
of $10^5$ four fermion events and show the effects on
the considered distributions of three different
selection criteria: fully extrapolated set-up (white histogram),
cuts on the scattering angles of $1,2$ particles coming from
the decay of the $W^-$ boson ($40^\circ \leq \vartheta_{1,2} \leq 140^\circ$)
(dark histogram)
and cut on the invariant mass of the $W^-$ boson
($77 \leq M_- \leq 83$~GeV) (grey histogram).
The effect of the cuts is shown for the $W^-$ angular distribution (Fig.~5),
the energy distribution of the fermion from $W^-$ decay (Fig.~6),
 the distribution of the relative angle of the decay products
of the $W^-$ boson (Fig.~7),
 the scattering angle of the fermion produced by $W^-$ decay (Fig.~8)
 and, finally, the photon energy distribution (Fig.~9).

\vskip 12pt\noindent
Acknowledegments -- We would like to thank the INFN, Sezioni di Pavia and
Turin,
for having provided computer resources. We are also grateful to Giorgio
Fumagalli and Valerio Vercesi
for helpful assistance in numerical and graphical problems.

\vfill\eject

\leftline{\large \bf Figure Captions}
\vskip 30pt
\noindent
Figure 1. The total cross section of $e^+ e^- \to W^+ W^- \to 4f$
without (dashed line) and with (solid line) initial state leading log
 QED corrections as a function of the c.m. energy and for
$W$'s decaying into all possible final states (top excluded). The open circles
are the results of our Monte Carlo program in comparison with the integration
of the analytic formula of ref.~[15] for the tree level approximation
and with the predictions of our semi-analytical code for the radiative case.

\vskip 8pt\noindent
Figure 2. The $W^-$ invariant mass distribution $d \sigma / d M_-$
($M_-^2 = - (q_1 + q_2)^2$) at the Born level for
 two c.m. energies  $\sqrt{s} = 175$~GeV
(dotted line) and $\sqrt{s} = 190$~GeV (dashed line).
The solid line is the QED-corrected invariant mass distribution at
 $\sqrt{s} = 175$~GeV. The Monte Carlo integration results are given by the
markers.

\vskip 8pt\noindent
Figure 3. The effect of
some experimental cuts on the
total cross section of $e^+ e^- \to W^+ W^- \to 4f$
 including initial state leading log
 QED corrections as a function of the c.m. energy, and for both
$W$ bosons decaying hadronically. The Monte Carlo integration results are
given by the markers.

\vskip 8pt\noindent
Figure 4. The same as in Fig.~3 for the invariant mass distribution at
$\sqrt{s} = 175$~GeV.

\vskip 8pt\noindent
Figure 5. The effect of specific selection criteria on the $W^-$ angular
distribution at $\sqrt{s} = 175$~GeV with QED corrections included:
no cuts (white histogram), cuts on the scattering
angles of $1,2$ particles coming from
the decay of the $W^-$ boson ($40^\circ \leq \vartheta_{1,2} \leq 140^\circ$)
(dark histogram)
and cut on the invariant mass of the $W^-$ boson
($77 \leq M_- \leq 83$~GeV) (grey histogram).

\vskip 8pt\noindent
Figure 6. The same as Fig.~5 for the fermion energy distribution.

\vskip 8pt\noindent
Figure 7. The same as Fig.~5 for the relative angle of the decay products
 of the $W^-$ boson.

\vskip 8pt\noindent
Figure 8. The same as Fig.~5 for the fermion scattering angle.

\vskip 8pt\noindent
Figure 9. The same as Fig.~5 for the photon energy distribution
($E_{\gamma} \leq 25$~GeV).

\end{document}